\documentclass[11pt,b5paper]{article}
\usepackage{natbib}
\usepackage[czech]{babel}
\usepackage[cp1250]{inputenc}
\usepackage{graphicx}

\newcommand{\oc}{\textit{O-C}}

\renewcommand{\title}[1]{\begin{center}{\large\bf #1}\end{center}}
\renewcommand{\author}[1]{\begin{center}#1\end{center}}
\newcommand{\zav}[1]{\left(#1\right)}
\newcommand{\hzav}[1]{\left[#1\right]}

\newcommand{\afill}[2]{\vspace{-18pt}\begin{center}\small\textit{#1}\\
\textit{#2}\end{center}}
\newcommand{\Section}[1]{\vspace{12pt}\par\noindent\stepcounter{section}
{\textbf{\thesection. #1}}\\[6pt]}
\renewcommand{\subsection}[1]{\vspace{6pt}\par\noindent\stepcounter{subsection}
\textit{\thesubsection. #1}\\[6pt]} \renewcommand{\caption}[1]{\vspace{6pt}
\par\noindent\refstepcounter{figure} {\small \textbf{Fig. \thefigure.} #1}}
\begin{document}
\thispagestyle{empty}
\title{TWO METHODS FOR THE LIGHT CURVES EXTREMA DETERMINATION}
\author{Z. Mikulášek$^{a,\,b}$ and T. Gráf$^a$ }
\afill{$^a$Observatory and Planetarium of J. Palisa, VŠB--Technical
    University of Ostrava, Czech Republic}{tomas.graf@vsb.cz}
\afill{$^b$Department of Theoretical Physics and Astrophysics,
    Masaryk University, Brno, Czech Republic}{mikulas@physics.muni.cz}
\begin{quotation}{\small\noindent \textbf{Abstract:} Two methods
for the determination of extrema timings and their uncertainties
appropriate for the analysis of time series of variable stars using
matrix calculus are presented. The method I is suitable for
determination of times of extrema of non-periodical variables or
objects, whose light curves vary. The method II is apt for \oc\
analyses of objects whose light curves are more or less repeating.
\\[6pt]
}
\end{quotation}
%
%

\Section{Linear regression using matrix calculus}

The determination of times of extrema and its uncertainty of light
variations belongs to the most common problems in the astrophysics of
variable stars. The following instructive text develops the basic
ideas outlined in \cite{mik} Mikulášek et al. (2006).

\subsection{The data}
Let we have $n$ observations $\{y_i\}$ creating the vector
$\mathbf{Y}$, $\mathbf{Y}=\{y_i\}$ done in moments $\{t_i\}$ creating
the vector $\mathbf{t}$, $\mathbf{t}=\{t_i\}$, both of the orders of
$n\times 1$. It is demanding to adapt the time scale so that
\textbf{t} $\Rightarrow$ $\mathbf{t}-\overline{\mathbf{t}}$.

The weight $w_i$ expresses the quality of an $i$-th measurement;
$\mathbf{w}=\{w_i\}$. If we know the inner uncertainty (error) of the
$i$-th observation $\delta y_i$, we can determine the weight of it as
follows: $w_i\sim(\delta y_i)^{-2}$. If we assume that the quality of
all measurements are more or less equal or if we know about the
quality of measurements little or nothing, it would be honest to put
$w_i\equiv 1$, $\mathbf{w}=\mathbf{E}(n,1)$.

The most of relations look simpler if we normalise weights so it is
valid $\overline{w}=1$, then $\sum_{i=1}^n w_i=n$. Frequently, we use
instead of the $\mathbf{w}$ square matrix ($n\times n$) $\mathbf{W}$,
$\mathbf{W}=\textrm{diag}(\mathbf{w})$.

\subsection{Linear regression}
The observed relationship between the dependent variable (inaccurately
measured quantity - mostly magnitude, radial velocity, temperature)
$y$ and the independent variable (precisely measured quantity --
typically time) $t$ can be fit by an appropriate \textbf{model
function} $F(t)$. The model function is determined by $g$ free
parameters $\beta_j$ that create a column vector
$\vec{\beta}=[\beta_1,\beta_2,...\beta_g]^\mathrm{T}$. The upper index
$^{\rm{T}}$ denotes transposing the matrix. If the model function
$F(t)$ can be expressed as a linear combination of $g$ different
functions of time $f_k(t)$ (we speak here about the linear model
function). Then
\begin{equation}
\mathbf{f}(t)=[f_1(t),\,f_2(t),...,f_g(t)], \quad F(t,\,
\vec{\beta})=\sum_{k=1}^g\,\beta_k\,f_k(t)=\mathbf{f}(t)\,
\vec{\beta}.
\end{equation}
Let introduce the matrix $\mathbf{X}$ of rank $n \times g$ and the
column vector $\mathbf{Y}_{\mathrm{p}},\,(n\times 1)$:
\begin{eqnarray}\label{X}
\mathbf{X}=\left(\begin{array}{c}
               \mathbf{f}(t_1) \\
               \mathbf{f}(t_2) \\
               ...  \\
               \mathbf{f}(t_n)\\
             \end{array}\right);
           \quad
\mathbf{Y}_{\mathrm{p}}=\left(\begin{array}{c}
               F(t_1) \\
               F(t_2) \\
               ...  \\
               F(t_n)\\
             \end{array}\right)=\mathbf{X\,\vec{\beta}}.
\end{eqnarray}
As the objective measure of the success rate of the fit for an a
$\vec{\beta}$ is used usually the sum of weighted squares of
deflection of observed values and predicted ones $S(\vec{\beta})$:
\begin{equation}
S(\vec{\beta})=(\mathbf{Y-Y}_{\mathrm{p}})^{\mathrm{T}}\,\mathbf{W}\,
(\mathbf{Y-Y}_{\mathrm{p}})=\mathbf{Y^{\mathrm{T}}\,W\,Y}
-2\,\beta^{\mathrm{T}}\,\mathbf{U}+\beta^{\mathrm{T}}\,\mathbf{V}\,\beta,\\
\end{equation}
where
\begin{equation} \mathbf{U}=\mathbf{X}^{\mathrm
T}\,\mathbf{W}\,\mathbf{Y};\quad \mathbf{V}=\mathbf{X}^{\mathrm
T}\,\mathbf{W}\,\mathbf{X};\quad \mathbf{H}=\mathbf{V}^{-1}.
\end{equation}
The least square method (LSM) considers the fit by the model function
$F(t,\vec{\beta})$ as the best one if the sum
$R=S(\vec{\beta}=\mathbf{b})$ is minimal. In the case of the linear
model function $F(t,\vec{\beta})$ we obtain for $\mathbf{b}$ and $R$:
\begin{equation} \label{minS}
\left. {\frac{\partial S}{\partial \vec{\beta}}}
\right|_{\vec{\beta}=\mathbf{b}}=\mathbf{0}=-2\,\mathbf{U}+2\,
{\mathbf{V}\,\mathbf{b}},\quad \Rightarrow\quad
\mathbf{b}=\mathbf{H}\,\mathbf{U};\quad R=\mathbf{Y}^{\rm
T}\mathbf{W}\,\mathbf{Y}-\mathbf{b}^{\rm{T}}\mathbf{U}.
\end{equation}

\subsection{The example}\label{exmpl}
Standardly used linear regression models are polynomials or
trigonometric functions of arbitrary orders. As an example we select
the parabolic model - the simplest model we can use for fit of the
real function in the form: $F(t)=\beta_1\,t^2+\beta_2\,t+\beta_3$,
$\mathbf{f}(t)=[t^2,\,t,\,1]$, $\mathbf{X}=[\{t_i^2\}\ \{t_i\}\ \{1\}
]$.

\subsection{Standard deviation. Uncertainties}
The standard deviation $\sigma$ can be estimated using relation
\begin{equation}
\sigma=\sqrt{\frac{R}{n-g}}.
\end{equation}
The components of the column vector $\delta \mathbf{b}$ used to be
considered as a rigourous estimate of the uncertainty of the
particular parameters. Unfortunately, they have this meaning only
exceptionally, nevertheless it is sometimes required by referees.
Contrary, very valuable is the following estimate of the uncertainty
of the model predictions $\delta F(t)$
\begin{equation}
\delta \mathbf{b}=\sigma\sqrt{\textrm{diag}(\mathbf{H})};\quad \delta
F(t)=\sigma\,\sqrt{\,\mathbf{f}(t)\,\mathbf{H}\,
\mathbf{f}(t)^{\rm{T}}}.
\end{equation}

\begin{figure}[h]
\centerline{\includegraphics[width=12cm,angle=0]{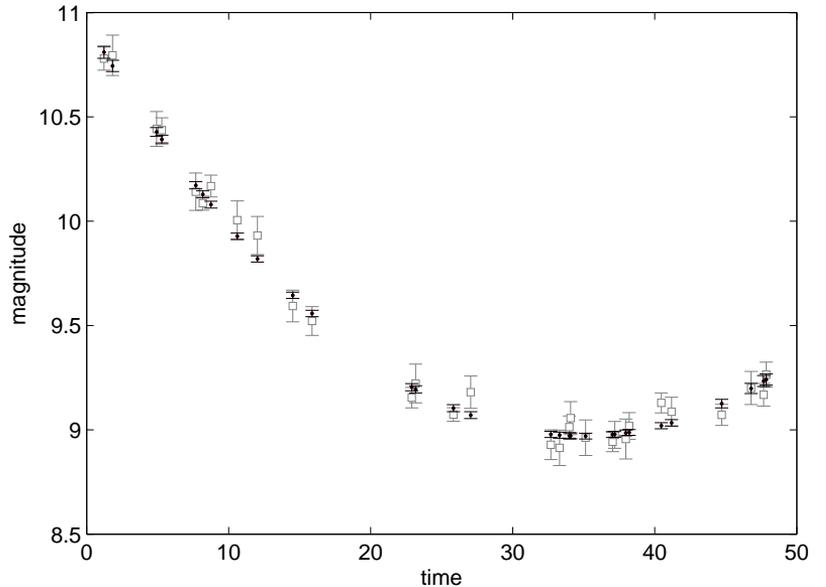}}
\begin{center}
\caption{\small The simulated light curve of a variable star with a
minimum. Inner accuracy of individual measurements are denoted by
gray error bars. The fitted parabola is marked by black dots with
error bars corresponding to the uncertainty of predicted values.}
\label{Fig1}
\end{center}
\end{figure}

\section{Method I}

Method I of the determination of the extrema times consists of
several steps.
\begin{itemize}
\item Plot the time series and select the appropriate linear model
function which can fit the observed light curve sufficiently well
with the minimum of free parameters.
\item Fit the observed light curve by the model function and check
whether the dependence of deflections of observed and predicted
light curves show only random scatter or some trends. In the latter
case improve your model function
\item Find the extreme time on the fitted light curve and calculate
its uncertainty.
\end{itemize}

\subsection{When the minimum/maximum occurs?}
The moments of the fitted model function extrema occurs if it is
fulfilled the condition that time derivative in that $t_{\rm{e}}$
equals to zero, especially
\begin{equation}\label{extreme}
0=\dot{y}_{\mathrm{p}}(t_{\rm{e}})=\dot{F}(t_{\rm{e}})=\frac{\mathrm{d}\zav{\mathbf{f}
(t_{\mathrm{e}})\,\mathbf{b}}}{\mathrm dt}=\dot{\mathbf{f}}
(t_{\mathrm{e}})\,\mathbf{b};\ \dot{\mathbf{f}} (t)=[\dot{f}_1(t),
\dot{f}_2(t),...\dot{f}_g(t)].
\end{equation}
There are many techniques how to find roots of the derivative of the
fitted function $\dot{F}(t)$. The uncertainty of the determination of
the time derivative at the time $t$, $\delta\dot{y}_{\mathrm{p}}(t)$
can be estimated by the relation:
\begin{equation}
\delta \dot{y}_{\rm{p}}(t)=\sigma\,\sqrt{\,
\dot{\mathbf{f}}(t)\,\mathbf{H}\,\dot{\mathbf{f}}(t)^{\mathrm{T}}}.
\end{equation}
For the estimate of the uncertainty of the time of extreme $\delta
t_{\rm{e}}$ we need yet the second time derivative in of the fitted
curve in the time of extreme $\ddot{y}_{\mathrm{p}}(t_{\rm{e}})$.
\begin{equation}
\ddot{y}_{\mathrm{p}}(t_{\rm{e}})=\ddot{\mathbf{f}}
(t_{\mathrm{e}})\,\mathbf{b};\quad \ddot{\mathbf{f}}
(t)=[\ddot{f}_1(t), \ddot{f}_2(t),...\ddot{f}_g(t)];\quad \delta
t_{\rm{e}}=\frac{\delta
\dot{y}_{\rm{p}}(t_{\rm{e}})}{|\ddot{y}_{\mathrm{p}}(t_{\rm{e}})|}.
\end{equation}

\section{Method II}

This method is again an application of the weighted least square
method this once with the model function defined so that time or times
of extrema are free parameters which are found and iteratively
improved. Uncertainties of these parameters are then simply
uncertainties of determined extrema moments.

The method is applied namely for periodic or nearly periodic
variable stars as pulsating stars, eclipsing binaries or rotating
stars. The disadvantage of the method consists in the fact that from
principle reasons we cannot applied here linear regression a we have
to used approaches established for solution of non-linear LSM.

\subsection{Example - transformation of the parabolic model}
In section 1.3 we introduced the linear model expression with three
parameters $\vec{\beta},\ F(t)=\beta_1\,t^2+\beta_2\,t+\beta_3$. This
parabolic model function can be rewritten in the non-linear form:
\begin{equation}\label{alpha}
F(t,\vec{\alpha})=\alpha_2\,(t-\alpha_1)^2+\alpha_3,
\end{equation}
where the parameter $\alpha_1=t_{\rm{min}}$ corresponds to the time of
the minimum of quadratic model function,
$\alpha_1=-\beta_2/2\,\beta_1,\ \alpha_2=\beta_1,\
\alpha_3=\beta_3-\beta_2^2/4\,\beta_1$. It is apparent that the model
$F(t,\vec{\alpha})$ as it is expressed in Eq.\,\ref{alpha} is no more
linear, nevertheless it is in the form demanded by the method II.

\subsection{Linearisation of non-linear model functions}
The solution of non linear regression is not straightforward and
immediate than in the case of linear regression. Nevertheless, having
a satisfactorily an initial estimate of the solution $\vec{a}_0$ we
can linearise the non-linear model function and delight in all
advantages yielding by linear models. The linearisation can be done by
the Taylor decomposition of the first order
\begin{equation}\label{decomp}
F(t,\vec{a_1})\cong F(t,\vec{a_0})+\sum_{j=1}^g\,\Delta
a_j\,\frac{\partial F} {\partial \alpha_j}
\end{equation}
The function is now linear in respect to new set parameters $\Delta
\mathbf{a}$, functions of the linearised model are
$\mathbf{f}(t)=[\frac{\partial F(t)} {\partial
\alpha_1},\frac{\partial F(t)} {\partial \alpha_2},...\frac{\partial
F(t)} {\partial \alpha_g}]$.

Now we can create matrices \textbf{X} (see Eq.\,\ref{X}), \textbf{Y},
\textbf{W} and calculate a solution for the vector $\Delta
\mathbf{a}$.
\begin{equation}
\mathbf{V=X^T\,W\,X;\quad U=X^T\,W\,X;\quad H=V^{-1};\quad \Delta
a=H\,U.}
\end{equation}
This correction we add to the initial estimate $\mathbf{a}_0$ and we
obtained the further, improved estimate for the vector of parameters
$\mathbf{a}_1=\mathbf{a}_0+\Delta \mathbf{a}$. The new value of
parameter set we can repeat the whole procedure. The iterative
process diminish the value of the correcting vector $\Delta
\mathbf{a}$ as rule very effectively and after several steps we
obtain the final result.

You need not to iterate all parameters, some of them (the linear
ones) can be calculate directly - see the following example.

\subsection{Example - linearisation of the parabolic model}
Now we can linearise our quadratic model according to the
Eq.\,\ref{decomp}, assuming the initial estimate of parameters
$\mathbf{a}_0$
\begin{equation}\label{alin}
F=\hzav{a_{02}\,(t-a_{01})^2+ a_{03}}+ \Delta a_1\,
2\,a_{02}\,(t-a_{01})+ \Delta a_2\,(t-a_{01})^2+\Delta a_3.
\end{equation}
The function is in parameters $a_2,\,a_3$ linear. It means we can
calculate the first two parameters directly, and $a_1$ iteratively:
\begin{equation}\label{alinda}
F(t,\mathbf{a})= \Delta a_1\ 2\,a_{2}\,(t-a_{01})+a_2\,(t-a_{01})^2+
a_3.
\end{equation}
\begin{eqnarray}
\mathbf{X}=\left(\begin{array}{ccc}
               2\,a_2\,(t_1-a_{01})& (t_1-a_{01})^2 &1   \\
               2\,a_2\,(t_2-a_{01})& (t_2-a_{01})^2 & 1  \\
               ... & ... & ...  \\
               2\,a_2\,(t_n-a_{01})& (t_n-a_{01})^2 & 1   \\
             \end{array}
           \right)
\end{eqnarray}
Knowing parameters $a_2,\,a_3$ (repeating light variations) we should
fix them.

\subsection{Estimation of the uncertainty of extrema timings}
The uncertainties of free parameters including the extremum timing are
given by
\begin{equation}
\Delta \mathbf{Y=Y}-\mathbf{F(\mathbf {t,a})};\quad R=\Delta
\mathbf{Y}^{\rm{T}}\mathbf{W}\,\Delta \mathbf{Y};\quad \delta
\mathbf{a}=\sigma\sqrt{\textrm{diag}(\mathbf{H})}.
\end{equation}
Anyhow the resulting sets of correcting parameters are practically
'pure zero', their uncertainties differ from zero a correspond to
uncertainties of the particular parameters. It enables to determine
the reliable estimate for errors in extrema time determinations.
%

\section{Discussion and conclusion}
It can be also proven that in the case of identical model the both
methods are interchangeable and yield the identical results.

\vspace{12pt} \noindent
\textbf{Acknowledgements}\\[6pt]
We thank M. Zejda for valuable discussion. This work was funded by the
following grants: GAAV IAA301630901, and MUNI/A/0968/2009.
\renewcommand{\refname}{
%
%
\end{document}